\documentstyle[draft,12pt]{article}

\begin{document}
\pagestyle{empty}
\newcommand{\dslash}{\mbox{ $\not \!\! D$}}

\newcommand{\unmezzo} {\frac{1}{2}}
\newcommand{\unterzo} {\frac{1}{3}}
\newcommand{\unquarto}{\frac{1}{4}}

\newcommand{\parzialet}[1]   {  \frac{ \partial{#1} }{\partial{t}} }
\newcommand{\parzialexi}[1]  {  \frac{ \partial{#1} }{\partial{x_i}} }
\newcommand{\parzialexxi}[1] {  \frac{ {\partial}^2{#1} }{\partial {x_i^2}} }
\newcommand{\derivatax}[1] {  \frac{        d{#1} }{        dx} }
\newcommand{\derivatat}[1] {  \frac{        d{#1} }{        dt} }

\newcommand{\traccia}{ {\rm Tr} }

\newcommand{\be}{\begin{equation}}
\newcommand{\ee}{\end{equation}}
\newcommand{\ca}{ {\cal A} }
\newcommand{\ch}{ {\cal H} }
\newcommand{\cm}{ {\cal M} }
\newcommand{\lam}{ {\lambda} }

\title{ \vspace{-2em} Fluid Random Surfaces with Extrinsic Curvature: II}
\author{
Konstantinos Anagnostopoulos, Mark Bowick, \\
Paul Coddington,  Marco Falcioni, Leping Han, \\
Geoffrey Harris and Enzo Marinari$^{(\dagger)}$ \\ [1em]
{ \normalsize Dept. of Physics and NPAC, Syracuse University,}\\
{ \normalsize Syracuse, NY 13244-1130, USA} \\ \\
{\footnotesize konstant bowick
paulc falcioni han gharris @npac.syr.edu marinari@roma1.infn.it}\\ \\
{\footnotesize $(\dagger)$:  and Dipartimento di Fisica and INFN,
Universit\`a di Roma {\it Tor Vergata}} \\
{\footnotesize Viale della Ricerca Scientifica, 00173 Roma, Italy}}

\date{\today}
\maketitle

\begin{center}
{\bf Abstract\vspace{-.2em}\vspace{0pt}}
\end{center}
{ \footnotesize
We present the results of an extension of our previous work on large-scale
simulations of dynamically triangulated toroidal random surfaces
embedded in $R^3$ with extrinsic curvature. We find that the
extrinsic-curvature specific heat peak ceases to grow on lattices with
more than $576$ nodes and that the location of the peak $\lam_c$
also stabilizes. The evidence for a true  crumpling transition is still
weak. If we assume it exists we can say that the finite-size scaling exponent
$\frac {\alpha} {\nu d}$ is very close to zero or negative.
On the other hand our new data does rule out the observed peak as being a
finite-size artifact
of the persistence length becoming comparable to the extent of the lattice.
}

\begin{flushright}
  {SU-HEP-93-4241-540}\\
  {SCCS-511}\\
  {hep-th/9308091}\\
\end{flushright}

\newpage

%\section{Introduction \protect\label{S_INT} }
\pagestyle{plain}
The theory of $2d$ fluid random surfaces embedded in $R^3$, with an extrinsic
curvature term (bending rigidity) in the action, has received considerable
analytical and numerical attention in the last decade
\cite{DAVREV,AMBTWO,BCHHM}.
In \cite {BCHHM} we presented the results of a large-scale Monte Carlo
simulation
of a dynamically-triangulated torus in $R^3$ with up to $576$ nodes,
corresponding to $1152$ triangles.
Although we observed a rapid crossover from a {\em crumpled} regime for
$\lam < \lam_c$
to a {\it smooth} regime for $\lam > \lam_c$, where $\lam$ is the extrinsic
curvature
coupling constant and $\lam_c \approx 1.425$, it was not at all clear whether
a true continuous thermodynamic phase transition separated the two regimes.
In fact several alternative interpretations of the data were discussed in
\cite {BCHHM}.
Perhaps the simplest possibility, advocated in \cite{KROGOM}, is that the
persistence length
$\xi$ describing the exponential decay of the normal-normal two-point function
in the crumpled (disordered) regime simply reaches the finite size of  the
system at $\lam_c$.
In this case the observed smooth regime would be a finite-size artifact with
the true continuum theory really being crumpled for all couplings $\lam$, in
accordance with
perturbative analytical results \cite{HELFRI,KLEINB,FORSTE,POLEXT}.
Since $\xi$ grows exponentially with $\lam$, according to the one-loop
beta-function,
this interpretation would imply that $\lam_c$ diverges logarithmically with
system size $N$.
To resolve this issue and to gain further insight into the model
it was clearly desirable to extend the numerical simulations to
larger lattice sizes and to clarify the influence of finite-size effects.
In this short letter we present an extension of our previous work to
include toroidal lattices  $1152$ and $2304$ nodes.

%\section{The Numerical Simulation \protect\label{S_NUM}}

As in \cite{BCHHM} we study the theory defined by the action

\be
	S = S_{Gauss} + \lam S_E =
     \sum_{i,j,\mu}C_{ij}(X^{\mu}_i - X^{\mu}_j)^2 +
\lam \sum_{\hat{k}\hat{l}}(1 - n^{\mu}_{\hat{k}}\cdot n^{\mu}_{\hat{l}})\  ,
\ee
where $C_{ij}$ is the adjacency matrix, $X^{\mu}_i$ is the position in $R^3$
of node $i$ ($i=1,..,N$)
and $n^{\mu}_{\hat{k}}$ is the normal vector to a triangle $\hat{k}$
in the cellular decomposition
of a lattice discretization of a torus.
The discretization $S_E$ of a continuum extrinsic curvature term
takes support on the edges (links) of the lattice and is known
as the (discrete) edge extrinsic-curvature.
The simulation consists of a standard Metropolis algorithm for the
updating of the nodes $X^{\mu}_i$ and a DTRS-algorithm \cite{EALRYM,MINOEX}
to sweep through the space of triangulations.
The basic flip move is attempted on randomly chosen links.
After a set of $3N$ flips are performed, $3N$ randomly selected
embedding coordinates are updated by random shifts from a flat
distribution.

The observable of most direct physical interest is the edge extrinsic-curvature
specific heat
\be
	C(\lam) = \frac{\lam^2} {N} (<S_E^2> - <S_E>^2)\ .
\ee
This exhibits a peak at a coupling $\lam_c$ which depends on the exact
discrete form of the action chosen
\cite{CATTER,BAJOWI,BCJW,CAKORE,CEKR,AMBONE,AMBTWO,BCHHM}.
In \cite{BCHHM} we found that the maximum value of the specific heat
grows with the system size as $C_{max} = A N^{\omega}$, with
$\omega = 0.06 \pm 0.05$.
In our new series of simulations on lattices with $1152$ nodes
we ran $54$ million sweeps at $\lam=1.425$, $21$ million sweeps at
$\lam=1.430$ and $18$ million sweeps at $\lam=1.435$
\footnote{ Since the autocorrelation time $\tau$ is of order $400,000$
sweeps on the $1152$ lattice these runs have at least $45 \tau$
measurements.}.
On the data from these three points we use multi-histogram reconstruction
\cite{ENZOHI,SWESTA}. This works well in that three different reconstructions
give coherent results.
On lattices of $2304$ nodes we have poorer statistics. We ran $17$ million
sweeps at $\lam=1.425$ plus approximately $5$ million sweeps at
$\lam = 1.40$, $1.42$ and $1.43$ as a consistency check.
On the $2304$ lattice histogramming does not work well.
This is to be expected since the statistics are not good enough
for such a large lattice. Still we have checked that our measurements
at $\lam=1.425$ give consistent results, that the error estimate is
reliable and that we are, with good accuracy, at the peak of the
specific heat.
In all these simulations required the equivalent of approximately one year
of CPU time on an HP 9000 (720 series) workstation.

The specific heat peak for $N=576$, $1152$ and $2304$ is shown in Fig.~1.
In Table~1 we give our results for the maximum of the specific
heat and the associated coupling $\lam_c$ as a function of $N$
\footnote{We have reanalyzed the data presented in reference
(\cite{BCHHM}), using a different method of weighting relative errors
when combining histograms.
Thus, some of the errors quoted here
are smaller than the respective uncertainties in reference (\cite{BCHHM}).}.

\begin{table}
\begin{center}
\begin{tabular}{||r||r|r||} \hline
$N$ & $C^{(max)}$ & $\lambda_c$ \\ \hline
 $  36$ & $3.484 \pm .008$ & $ 1.425 \pm .035$ \\ \hline
 $  72$ & $4.571 \pm .015$ & $ 1.410 \pm .015$ \\ \hline
 $ 144$ & $5.37  \pm .08 $ & $ 1.395 \pm .017$ \\ \hline
 $ 288$ & $5.55  \pm .05 $ & $ 1.410 \pm .015$ \\ \hline
 $ 576$ & $5.81  \pm .06 $ & $ 1.425 \pm .010$ \\ \hline
 $1152$ & $5.69  \pm .04 $ & $ 1.425 \pm .010$ \\ \hline
 $2304$ & $5.75  \pm .10 $ & $ 1.425 \pm .010$ \\ \hline
\hline
\end{tabular}
\end{center}
\protect\caption[CT_ONE]{The maximum of the specific heat and its position,
with errors, for different lattice sizes.
\protect\label{T_ONE}}
\end{table}

Clearly the maximum of the specific heat curve $C_{max}$ is effectively
constant for surfaces with $576$ or more nodes.
The (pseudo)-critical coupling $\lam_c$ is also constant for $N=576$
and above.
With the present data we can definitely exclude the presence of a divergence
in the specific heat. The growth of the specific heat peak observed on small
lattices \cite{CATTER,BAJOWI,BCJW} does not reflect true asymptotic behaviour.
These results also invalidate the interpretation raised in the
introduction \cite{KROGOM,BCHHM}.
A one-loop renormalization group calculation shows that the persistence
length $\xi$ grows with bending rigidity $\lam$ as
${\rm exp}({\frac {3} {4\pi}} \lam)$.
Equating $\xi$ with the spatial extent of the lattice $N^{1/d_{in}}$,
where $d_{in}$ is the intrinsic Hausdorff dimension of the lattice,
one sees that they become comparable at a coupling
$\lam_c \sim {\frac {3} {4\pi d_{in}}} {\rm ln} (N)$.
In the continuum limit $N \rightarrow \infty$, $\lam_c$ diverges.
Since, for reasonable values of $d_{in}$, we do not see the increase
in $\lam_c$ with $N$ predicted by the above relationship we can state
with some confidence
that the origin of the observed specific heat peak is not explained
by the persistence length becoming comparable to the extent of the
lattice.

In \cite{BCHHM} we also discussed other, more subtle, possibilities
that could account for the observed behaviour of $C(\lam)$ without
invoking a phase transition.
One was based on the analogy between the present model and the $O(3)$
sigma-model in $2d$ \cite{ZINJUS}.
This model is also asymptotically free
and consequently disordered at all non-zero temperatures.
Yet numerical simulations show a distinct peak in the specific heat
which grows for small lattices and then saturates, just as we find
in the model of a rigid string treated here.
This may be seen in Fig.~2 where we have plotted the specific heat
$C(\beta)$ for the two dimensional $O(3)$ model. The simulations
were done  on  square
lattices of volume $N=16$, $25$, $64$, $100$, $900$, $2,500$,
$4,900$ and $10,000$
using the Wolff algorithm \cite{WOLFFO}. For each point of the
$N=25,100$ and $10,000$ lattices we used $100,000$
measurements. We took a measurement every time the Wolff clusters
updated a  volume exceeding $30$
times the volume of the lattice
\footnote{For the $N=16$, $64$, $900$ and $4,900$ lattices the integrated
autocorrelation times were between $1$ and $2$ Wolff updatings
of the entire lattice}.
For each point of the $N=16$, $64$, $900$, $2,500$ and
$4,900$ lattices we used $20,000$ measurements.
We took a measurement every time
the Wolff clusters updated a volume exceeding $3$ times the volume of the
lattice.
It is very clear that the peak
levels off quickly for $N\ge 100$ and that ``$\beta_c$'' is not
increasing with the size of the lattice.  Measurements of the
asymptotic value of $C(\beta)$ have been reported in the past
\cite{MAPAPE,COLOT}. The authors of
\cite{BRDE,BRDESI,ORBR} explain the peak as the excitation of an
extra degree of freedom, the so-called $\sigma$-particle
\cite{BALESH}. The would-be  transition occurs when the mass of the
$\sigma$-particle becomes comparable to the inverse correlation length
of the $O(3)$ model. It may be that there is a similar
interpretation of the observed peak of $C(\lambda)$.

We are currently histogramming our data to examine as well the behaviour
of the complex zeroes of the partition function when  $\lam$ is
allowed to become complex. For $SU(2)$ lattice gauge theory, which also
exhibits a specific heat peak without an associated phase transition,
it has been shown that there are complex zeroes which are near the real axis
but do not converge to it in the infinite-volume limit \cite{ENZOHI}.
High-temperature expansions also indicate that the $O(3)$ model
susceptibility has a complex singularity near the real axis \cite{BUTERA}.

It is still possible that there is a true continuous phase transition,
the crumpling transition, occurring at $\lam_c$. Assuming a continuous
transition, a standard finite-size-scaling argument only tells us that
$\omega= \frac{\alpha} {\nu d} < 0$, where $\alpha$ is the exponent governing
the divergence of the specific heat,
$C(\lam) \sim {\vert\lam - \lam_c\vert}^{-\alpha}$,
$\nu$ is the analogous exponent for the
correlation length and $d$ is the intrinsic Hausdorff dimension of the
surface.  In other words there may be a cusp singularity at $\lam_c$
as, for example, in the case of the superfluid ($\lambda$) transition
in $He^4$ \cite{LIPA,GOLBOOK}, for which $\alpha = -0.0127 \pm 0.0026$.
Since we do not have a measurement of $\nu d$, which may even be
rather large, we have no reliable idea of the exponent $\alpha$ itself.
Generally speaking one finds that second order transitions
on fixed lattices become higher order on dynamical lattices,
as for example in the case of the $2d$-Ising model \cite {KAZIS,BOUKAZIS}.
Since there seems to be a $2$nd order  crumpling
transition for non-self-avoiding tethered (fixed-triangulation) surfaces
\cite{KANNEL,AMDUJO,WHEATE,WHESTE},
it would be consistent for the transition to be
higher than $2$nd order when the model is coupled to gravity.

All told our work gives only weak evidence for a continuum
crumpling transition. The strongest evidence in favour of
such a transition, at present, is the scaling behaviour of the
string tension and mass gap reported in \cite{AMBTWO}.
This highlights the need for more extensive measurements on these
important observables.

We have also measured the fluctuations of the extrinsic Gaussian
curvature $|{\cal K}|$, defined as the average magnitude of
the deficit angle at each vertex as measured in the embedded space.
Likewise, we have also computed the fluctuations of the mean
defect coordination number $|q - 6|$,
which is proportional to the intrinsic Gaussian curvature.
In reference \cite{BCHHM} we had observed
that fluctuations of these observables were quite large near the coupling
$\lambda_c$ but then drop
quite dramatically for slightly higher $\lambda$.
%rough transient region that we were not able to
%understand in detail.
We find that on larger lattices the fluctuations
of these observables at $\lambda_c$ also do not grow with $N$;
thus their behaviour does not provide unequivocal evidence of
the presence of a phase transition.
%This seems to have disappeared on the larger
%lattices and we do not see any sign of an impending
%divergence.
In Table~2 we give the mean-square fluctuations
of both observables.

\begin{table}
\begin{center}
\begin{tabular}{||r||r|r||} \hline
$N$ & $F[K]$ & $F[q-6]$ \\ \hline
 $ 576$ & $5.71  \pm .08 $ & $ 8.39 \pm .04$ \\ \hline
 $1152$ & $5.59  \pm .05 $ & $ 8.32 \pm .03$ \\ \hline
 $2304$ & $5.70  \pm .10 $ & $ 8.37 \pm .06$ \\ \hline
\hline
\end{tabular}
\end{center}
\protect\caption[CT_TWO]{The mean square fluctuations of the extrinsic
Gaussian curvature ${\cal K}$ and the defect coordination number $q-6$,
with errors, for different lattice sizes.
\protect\label{T_TWO}}
\end{table}

Finally we note the behaviour of the gyration radius at $\lambda_c$
(which we will take as being $1.425$ in the following). For large
$\lambda$ ($>2$),  the scaling of $R(N)
\sim N^{\frac{2}{d_H}}$
with $N$ gives a Hausdorff dimension close to $2$ (as we expect
for flat surfaces). In the crumpled region the Hausdorff dimension
rapidly increases with diminishing $\lambda$.
We had pointed out in \cite{BCHHM} that finite size effects
were relevant in the sector close to $\lambda_c$ and that we could not
estimate a reliable number from the lattice sizes analyzed. Here the
largest lattice we simulated ($N=2304$) does not give useful
data, since the error in $R$ is too large, but on the $1152$ and $576$
node lattices we get a fairly precise estimate of $R$, which allows us to
estimate for the Hausdorff dimension at the pseudo-critical point
$\lambda_c$ the value $d_H = 4.35 \pm .3$.
This is an intriguing result, since
$4$ is the extrinsic Hausdorff dimension of a class of branched
polymers, as constructed, for instance, in references {\cite{AMBBP}}.
Such configurations are expected to dominate the string functional
integral for large embedding dimension $D$.

%\section*{Acknowledgments}

This work has been done with NPAC (Northeast Parallel Architectures Center)
and CRPC (Center for Research in Parallel Computing) computing facilities.
The research of MB was supported by the Department of Energy Outstanding
Junior Investigator Grant DOE DE-FG02-85ER40231 and that of GH by research
funds from Syracuse University. KA wishes to thank John Apostolakis for
discussions and for providing him with his $O(3)$ code and Tryphon
Anagnostopoulos  and Alexis Arvilias for hospitality at the Democritos
Nuclear Center in Athens, Greece where part of this work was completed.
LH was supported by NPAC research funds.
We gratefully acknowledge discussions with Jan Ambj{\o}rn,
Geoffrey Fox, David Nelson and Bengt Petersson.

\vfill

\newpage

\section{Figure Captions}
  \begin{itemize}

    \item[Fig. 1] The edge extrinsic-curvature specific heat
$C(\lam)$ as a function of
$\lambda$. Multi-histogram reconstructions with errors are shown for $N=576$
(long and short-dashed lines) and $N=1,152$ (solid lines).
Four individual data points are also shown for $N=2,304$ (solid circles).
One sees that the specific heat peak has saturated - it is not growing
with the system size $N$ above $576$.

    \item[Fig. 2] The specific heat $C(\beta)$  of
the two-dimensional $O(3)$ non-linear sigma model as a function of
$\beta$ for lattice volumes $N=16$, $25$, $64$, $100$, $900$, $2,500$,
$4,900$ and $10,000$. The peak saturates
quickly for $N\ge 100$ and ``$\beta_c$'' does not increase with the volume.

  \end{itemize}

  \vfill

\end{document}